\documentclass{article}
\usepackage{amsthm,amsmath,amssymb,amsfonts,graphicx,subfigure,bm}  
\usepackage{makecell}
\usepackage{caption}
\usepackage{geometry}
\usepackage{tikz}
\usepackage{color}
\usepackage[OT1]{fontenc}
\usepackage[utf8]{inputenc}
\usepackage{graphicx,booktabs}
\graphicspath{{./figures/}}
\usepackage{threeparttable}
\usepackage{hyperref, cleveref}
\usepackage{autonum}
\usepackage{multirow}
\usepackage{hyperref}
\usepackage[section]{placeins}
\usepackage{soul}
\usepackage{tabularx}
\usepackage{float}
\usepackage{rotating}
\usepackage{lscape}
\usepackage{bbm}
\usepackage{natbib}
\DeclareMathOperator*{\argmax}{argmax}
\DeclareMathOperator*{\argmin}{argmin}
\newcommand{\BMA}{BayesMAR-BMA}
\newcommand{\arBMA}{BayesAR-BMA}
\newcommand{\MAP}{BayesMAR-MAP}
\newcommand{\arMAP}{BayesAR-MAP}

\newcommand{\Laplace}{\mathrm{Laplace}}

\usepackage{lineno}

\begin{document}

\title{Bayesian Median Autoregression for Robust Time Series Forecasting}
\author{Zijian Zeng and Meng Li \\ Department of Statistics, Rice University}
\maketitle

\begin{abstract}
	We develop a Bayesian median autoregressive (BayesMAR) model for time series forecasting. The proposed method utilizes time-varying quantile regression at the median, favorably inheriting the robustness of median regression in contrast to the widely used mean-based methods. Motivated by a working Laplace likelihood approach in Bayesian quantile regression, BayesMAR adopts a parametric model bearing the same structure of autoregressive models by altering the Gaussian error to Laplace, leading to a simple, robust, and interpretable modeling strategy for time series forecasting. We estimate model parameters by Markov chain Monte Carlo. Bayesian model averaging is used to account for model uncertainty including the uncertainty in the autoregressive order, in addition to a Bayesian model selection approach. The proposed methods are illustrated using simulations and real data applications. An application to U.S. macroeconomic data forecasting shows that BayesMAR leads to favorable and often superior predictive performance compared to the selected mean-based alternatives under various loss functions that encompass both point and probabilistic forecasts.  The proposed methods are generic and can be used to complement a rich class of methods that build on the autoregressive models.  
\end{abstract}

\section{Introduction}
\label{Section:Intro}

Time series forecasting is a long-standing problem in econometrics and statistics, where the overwhelming focus has been on mean-based models~\citep{prado2010time, hyndman2018}. Although conditional means are the optimal forecast under the squared error loss, complex characteristics violating model assumptions that are present in real data may hamper the predictive performance. Flexible nonparametric methods~\citep{ferraty2006nonparametric, fan2008nonlinear} building on minimal assumptions are an appealing remedy; however, they often have disadvantages in terms of not only interpretability but also in involving large numbers of parameters that often lead to daunting computation and communication issues. We were thus motivated by an attempt to propose a simple, interpretable, and principled strategy that can improve upon mean-based models in real-world time series forecasting. 

There is a rich literature on robust time series forecasting including categorizing outliers~\citep{Fox1972, HR2014}, adjusting autoregressive (AR) models to offset effects of outliers~\citep{CL1993, CLJ1993}, exponential smoothing and Holt-Winters seasonal methods to M-estimation~\citep{CGF2008}, weighted forecasts~\citep{JW2008}, and detecting structural changes~\citep{Q2008, OQ2011, CP2019}, just to name a few. As a nonlinear alternative, the median is more robust than the mean. While the application of median-based methods in time series at least dates back to 1974 when John Tukey introduced the running median method~\citep{JWT1974}, there is surprisingly little work to comprehensively investigate the modeling, fitting, and uncertainty quantification of a median-based model in the context of time series forecasting, and in particular, how it compares with state-of-the-art mean-based methods using real data and under various loss functions. 

In the quantile regression literature that encompasses the median as a special case~\citep{KB1978}, \cite{KX2006} proposes quantile autoregression (QAR) models which depict the conditional distributions of the response more comprehensively at various quantile levels.~\cite{EM2004} proposed the conditional autoregressive value at risk (CAViaR) model for risk management at extreme quantile levels, and this model has been followed by many others \citep{GK2005, CS2006, KG2007, GCC2011}. However, they were developed either for general quantile levels or extreme quantile levels and have not been applied to time series forecasting. Moreover, QAR and CAViaR are semiparametric approaches, resorting to minimizing the check loss function for estimation and necessitating non-trivial modifications to likelihood-based order selection criteria when the order is unknown. 

In this paper, we propose a simple strategy by extending the traditional AR model to a median AR model (MAR) for time series forecasting. The AR model is arguably one of the most popular methods in time series, serving as the building block for other models such as generalized autoregressive conditional heteroskedasticity models \citep[GARCH,][]{bollerslev1986} and time-varying vector autoregressive models \citep[TV-VAR][]{primiceri2005time}. The proposed method utilizes time-varying quantile regression but focuses on the median, favorably inheriting the robustness of median regression in contrast to the widely used mean-based methods. It relies on parametric assumptions, and this aids interpretation and enables convenient uncertainty quantification and propagation through a principled Bayesian framework. Numerical experiments using U.S. macroeconomic data show that this simple MAR approach leads to favorable and often superior predictive performances compared to selected state-of-the-art mean-based methods that are much more complicated in nature. It is remarkable that the comparison is made not only under the absolute but also the squared error loss for point forecasts and is extended to probablistic forecasts. The proposed methods are generic and can be used to complement any methods that build on the AR models by altering the Gaussian error assumption therein. 

The MAR model has a close connection with the working asymmetric Laplace likelihood approach in Bayesian quantile regression. The working likelihood formalized by~\cite{YM2001} has recently gained increasing attention~\citep{GCC2011, SK2016, BD2018, LLM2018}. It provides a principled and convenient framework to quantify uncertainties in the parameter estimation eliminating the challenging task of estimating the unknown conditional density functions that are required by the conventional quantile regression for inference~\citep{yang2016posterior}. Although the asymmetric Laplace likelihood is generally not the true data generating likelihood, a pragmatic view to support its use is that the maximum a posterior estimates resemble the usual quantile regression estimates which optimizes the check loss function. Theoretically, the posterior distribution of parameters concentrates on what minimizes the Kullback-Leibler divergence with respect to the true data-generating models~\citep{kleijn2006}. Unlike quantile regression where one may vary the quantile levels, the median is the primary quantile level of interest in time series forecasting. Therefore, the MAR model uses the Laplace distribution as the likelihood, alleviating the concern that the working likelihood is not a valid likelihood when considering multiple quantile levels. This fully parametric model enables routine posterior sampling. We estimate model parameters by Markov chain Monte Carlo, and propose using a Bayesian model averaging (BMA) approach~\citep{hoeting1999} to propagate the model uncertainty in the unknown autoregressive order in addition to a Bayesian model selection strategy.

The rest of the paper is organized as follows. Section~\ref{sec:method} introduces the MAR model, estimation, and forecasting procedure. In Section~\ref{sec:simulation} we conduct simulations to compare the proposed approach with other competitive methods to assess parameter estimation. Section~\ref{sec:Real} consists of a variety of applications using real-world economic data. Section~\ref{sec:conclusion} concludes the paper.

\section{Methods}
\label{sec:method}
\subsection{Median Autoregressive (MAR) model}
\label{subsec:2.1}
We propose a median autoregressive (MAR) model for time series forecasting, as an alternative to the widely used mean-based models. Suppose we observe a vector of time series $\bm{y} = (y_1, \ldots, y_T)$. A MAR model with order $p$, denoted as MAR$(p)$, first assumes a time-varying quantile regression structure 
\begin{equation} \label{eq:model.semiparametric} 
	y_t  =  \beta_0 + \beta_1 y_{t-1} + \beta_2 y_{t-2} + ... + \beta_p y_{t-p} + \epsilon_t = \bm{y}_{t-1}'\bm{\beta} + \epsilon_t,
\end{equation}
where $\bm{y}_{t-1} = (1, y_{t-1}, ..., y_{t-p})'$,  $\bm{\beta} = (\beta_0, \beta_1, ..., \beta_p)^\prime$ is vector of unknown coefficient, and $\epsilon_t$ is random error with \textit{median} 0.  Model~\eqref{eq:model.semiparametric} is \textit{semiparametric} as the error distribution is left unspecified other than the constraint of possessing a zero median. The classic AR model with a given order $p$, denoted as AR$(p)$, assumes a Gaussian error distribution with mean 0 and standard deviation $\sigma$, i.e., $\epsilon_t \sim N(0, \sigma^2).$ The MAR model further assumes $\epsilon_t \sim \Laplace(0, 2\tau)$ whose probability density function is 
\begin{equation} \label{eq:model.error} 
    f(x; 0, \tau) = \frac{1}{4 \tau} \exp\left(-\frac{|x|}{2\tau}\right), 
\end{equation}
where $\tau>0$ is a scale parameter. The Laplace error assumption combined with the semiparametric structure in Equation~\eqref{eq:model.semiparametric} yielding the following likelihood function for the MAR model
\begin{eqnarray} \label{eq:likelihood}
L(\bm{\beta}, \tau) = f(\bm{y} | \bm{\beta}, \tau)  \propto \tau^{-(T-p)} \exp\left\{ -\frac{1}{2\tau}\sum^{T}_{t=p+1} |y_t -\bm{y}_{t-1}' \bm{\beta} | \right\},
\end{eqnarray} 
which is \textit{parametric}. The parametric assumption in~\eqref{eq:model.error} aids interpretation and enables convenient uncertainty quantification and propagation through a principled Bayesian framework. In addition, the autoregressive structure in the MAR model resembles the widely used AR model provides enormous flexibility and potential to complement the rich literature that builds on the AR model. 

The use of Laplace distributions is common in literature the Bayesian quantile regression where the $\theta$th quantile of the error in Equation~\eqref{eq:model.semiparametric} is assumed to be 0.  For general $\theta \in (0, 1)$, a working likelihood method adopts the asymmetric Laplace distribution $\mathrm{AL}(\mu, \tau, \theta)$ as the error distribution, which has the probability density function  
\begin{equation}
	f(x; \mu, \tau, \theta) = \theta(1-\theta)\tau^{-1} \exp \left\{- \tau^{-1}(x - \mu)( \theta - \mathbbm{1}[x<\mu]) \right\},
\end{equation} 
where $\mu$ is a location parameter and $\mathbbm{1}(\cdot)$ is the indicator function. The asymmetric Laplace distribution reduces to the Laplace distribution at the median by setting $\mu = 0$ and $\theta = 0.5$. In view of this intimate connection with Bayesian quantile regression as well as the subsequent Bayesian estimation and prediction, we also refer to the MAR model with Laplace errors given by~\eqref{eq:model.semiparametric} and~\eqref{eq:model.error} as Bayesian MAR, or BayesMAR, and use it exchangeably with the MAR model.

\subsection{Prior specification and posterior sampling at a given $p$}
We use an uninformative prior for $\bm{\beta}$ and the Jeffreys prior for $\tau$, namely, 
\begin{equation}
	\pi(\bm{\beta})\propto 1, \quad \pi(\tau) \propto \tau^{-1}. 
\end{equation}
The posterior distribution of $(\bm{\beta}, \tau)$ is 
\begin{equation}\label{eq:posterior.beta.tau} 
\pi(\bm{\beta}, \tau | \bm{y}) \propto \pi(\bm{\beta},\tau) f(\bm{y} | \bm{\beta}, \tau) \propto \tau^{-(T-p+1)} \exp\left\{ -\tau^{-1}\sum^T_{t=p+1}\frac{1}{2}  \left| y_t - \bm{y}^\prime_{t-1}\bm{\beta} \right|\right\}, 
\end{equation}
which is proper~\citep{CH2013}. The regression coefficients $\bm{\beta}$ are instrumental for time series forecasting, and we derive their marginal posterior distributions by integrating out $\tau$ for efficient sampling: 
\begin{eqnarray}
	\pi(\bm{\beta} | \bm{y}) & = & \int{\pi(\bm{\beta}, \tau | \bm{y})} d\tau \propto \int \tau^{-(T-p+1)} \exp\left\{ -\tau^{-1}\sum^T_{t=p+1}\frac{1}{2}  \left| y_t - \bm{y}^\prime_{t-1}\bm{\beta} \right|\right\} d\tau \nonumber \\
	& \propto & \left\{ \sum^{T}_{t=p+1} \frac{1}{2} \left| y_t - \bm{y}^\prime_{t-1} \bm{\beta} \right|\right\}^{-(T-p)}. 
\end{eqnarray}
The posterior sampling of $\bm{\beta}$ proceeds by Markov chain Monte Carlo (MCMC) via the Metropolis-Hastings (MH) algorithm~\citep{MRRT1953, H1970}: 
\begin{enumerate}
	\item At iteration $i$, draw a candidate sample from the proposal $\beta^\ast_j = \beta^{i-1}_{j} + a\cdot u_j$ independently for $j = 1,\ldots, p$, where $\beta^{i-1}_{j}$ is the value of $\beta_j$ at iteration $i-1$, $u_j$ follows a Uniform$(-0.1, 0.1)$ distribution and the scalar $a$ controls the step size of each move; 
	\item Accept $\bm{\beta}^\ast$ as ${\bm{\beta}^i}$ with probability $p = \min\left\{1, \pi({\bm{\beta}}^\ast|\bm{y}) / \pi({\bm{\beta}^{i-1}}|\bm{y}) \right\}$. Otherwise, set ${\bm{\beta}^i} = {\bm{\beta}^{i-1}}$. 
\end{enumerate}
We tune the parameter $a$ in Step 1 such that the final acceptance rate is between $20\%$ and $50\%$~\citep{GRG1996}. In addition, we have also implemented Gaussian and heavy-tailed student-$t$ proposals, which are recommended by~\cite{GCC2011} when studying extreme quantile levels under the CAViaR model. We did not observe empirical advantages of using such proposals over a uniform proposal under the MAR model, suggesting that one may choose more flexible proposals for median regression. For all experiments, we use 40,000 MCMC samples with 25,000 burn-ins, initialize $\bm{\beta}$ randomly within the unit interval, and use the posterior mean as the Bayes estimate of ${\bm{\beta}}$. Trace plots indicate the MCMC samples converge quickly, mostly within thousands of iterations.

The proposed Bayesian approach provides a convinient method for density forecasting beyond point forecasts. The $1$-step ahead predictive density conditional on $\bm{y}_{1:T} = (y_1, \ldots, y_T)$ is
\begin{eqnarray*}
	p\left(y_{T+1} | \bm{y}_{1:T}\right) & = & \int \int p\left(y_{T+1}, \bm{\beta}, \tau | \bm{y}_{1:T}\right) d\bm{\beta} d\tau  \\
	& = & \int \int  p\left(y_{T+1} | \bm{\beta}, \tau  , \bm{y}_{1:T}\right)\cdot \pi\left( \bm{\beta}, \tau | \bm{y}_{1:T}\right) d\bm{\beta} d\tau \\
	& \propto & \int \int \tau^{-1} \exp \left( -\frac{1}{2\tau} |y_{T+1} - \bm{y}'_{T} \bm{\beta} | \right)\cdot \tau^{-(T-p+1)}\exp\left\{-\tau^{-1}\sum^T_{t=p+1} \frac{1}{2} |y_t - \bm{y}'_{t-1}\bm{\beta} |\right\} d\bm{\beta}d\tau \\
	& \propto & \int_{\mathbb{R}^{p+1}} \left[ \sum^{T+1}_{t=p+1} |y_t - \bm{y}'_{t-1}\bm{\beta} | \right]^{-(T-p+1)} d\bm{\beta}. \label{eq:closed.form}
\end{eqnarray*}
For large $T$ the integrand $\left[ \sum^{T}_{t=p+1} |y_t - \bm{y}'_{t-1}\bm{\beta}| \right]^{-(T-p+1)}$ may quickly decay to zero, leading to considerable numerical errors in direct evaluation of this integral. Alternatively, the MCMC samples of $(\bm{\beta}, \tau)$ allow a convenient sampling strategy to approximate the predictive density. In particular, we draw samples of $y_{T + 1}$ from the Laplace likelihood $p\left(y_{T+1} | \bm{\beta}, \tau  , \bm{y}_{1:T}\right)$ conditional on each posterior sample of $(\bm{\beta}, \tau)$. This strategy easily generalizes to $q$-step ahead predictive densities for any $q \geq 2$ by drawing samples jointly for $(y_{T + 1}, \ldots, y_{T + q})$ through iterative conditional distributions, which are all Laplace distributions.

\subsection{Order selection and Bayesian model averaging}
\label{sec:BIC}
The order $p$ in MAR($p$) is typically unknown. We address the problem of unknown $p$ in the Bayesian framework by putting a prior on $p$. In practice, we can usually specify a maximum order; otherwise, a $p$ that is too large hampers the interpretability. We endow the order $p$ with a uniform prior on $\{1, 2, \ldots, K\}$ with $K$ being the specified maximum order. Then the posterior distribution of $p$ in the prior support is 
\begin{align}
    \pi(p \mid \bm{y}) & \propto \pi(\bm{y} \mid p) \cdot \pi(p), 
\end{align}
where $ \pi(\bm{y} \mid p) = \int_{\mathbb{R}^+} \int_{\mathbb{R}^{p+1}} \pi(\bm{y}, \bm{\beta}, \tau \mid p) \pi(\bm{\beta}) \pi (\tau) d \bm{\beta} d{\tau}$ is the marginal likelihood of $p$. The order $p$ can be selected by using the maximum a posteriori (MAP) estimate: 
\begin{equation}
\hat{p} = \argmax_{p \in \{1, \ldots, K\}} \pi(p \mid \bm{y}). 
\end{equation}
For time series forecasting, a more appealing perspective is to use Bayesian model averaging (BMA) to propagate uncertainties in the model space, i.e., 
\begin{equation}
    \hat{y}_{T+q} = \sum_{p = 1}^K \pi(p \mid \bm{y}) \hat{y}^{(p)}_{T+q} \quad \text{and}  \quad p(y_{T + q} | \bm{y}_{1:T})= \sum_{p = 1}^K \pi(p \mid \bm{y})  p(y_{T + q} ^{(p)} | \bm{y}_{1:T}), 
\end{equation}
where $\hat{y}^{(p)}_{T+q}$ and $p(y_{T + q}^{(p)} | \bm{y}_{1:T})$ are the $q$-step ahead point prediction and predictive density of $y_{T + q}$ under order $p$, respectively. 

The main challenge in implementing MAP and BMA lies in prior specifications of model parameters and the evaluation of the marginal likelihood at given $p$. To date, a consensus on the default choice for prior specifications in the context of model selection is still lacking, and one needs to be cautious about using improper priors---which are typically the default choice for a given model---in view of prior sensitivity and the Jeffreys-Lindley paradox. We refer interested readers to the rich literature on BMA, e.g., \cite{cp2014, R2001, YVSG2018, LD2020}, to name a few. We here resort to an approximation to $\pi(p \mid \bm{y})$ using the Bayesian information criterion or BIC~\citep{RA1995,AJ2012}, which is appealing as it eliminates the need to deal with prior specification and approximates Bayes factors reasonably well in certain cases \citep{KW1995}. We observe that the BIC-based implementation tends to choose the oracle order with large probability in simulations. 

Letting $(\hat{\bm{\beta}}_{\text{MLE}}, \hat{\tau}_{\text{MLE}} )$ be the maximum likelihood estimates (MLEs) of $(\bm{\beta}, \tau)$, then the the BIC of MAR($p$) is
\begin{equation}
\label{eq:BIC.p} 
\text{BIC}_p =  (p+2)\log(n) - 2\log(L(\hat{\bm{\beta}}_{\text{MLE}}, \hat{\tau}_{\text{MLE}} )), 
\end{equation}
where $n$ is the sample size. We approximate $\pi(p \mid \bm{y})$ by $\exp\{-\text{BIC}_p/2\}$ up to multiplicative constants, leading to aggregated predictions 
\begin{eqnarray}
\label{eq:BMA.BIC}
\hat{y} = \sum_{p = 1}^K \pi(p \mid \bm{y}) \hat{y}^{(p)} \approx \sum_{p = 1}^K \frac{\exp\{-\text{BIC}_p /2\}}{\sum_{i=1}^K \exp\{-\text{BIC}_i /2\}} \hat{y}^{(p)}. 
\end{eqnarray} 
It turns out that we can calculate the MLEs $(\hat{\bm{\beta}}_{\text{MLE}}, \hat{\tau}_{\text{MLE}} )$ efficiently. To see this, first note 
\begin{equation}\label{eq:MAP.joint} 
(\hat{\bm{\beta}}_{\text{MLE}}, \hat{\tau}_{\text{MLE}} ) = \argmax_{\bm{\beta}, \bm{\tau}} \tau^{-(T-p)} \exp\left\{ -\tau^{-1}\sum^T_{t=p+1}\frac{1}{2}  \left| y_t - \bm{y}^\prime_{t-1}\bm{\beta} \right|\right\}.
\end{equation} For any $\tau > 0$, the likelihood function $L(\bm{\beta}, \bm{\tau})$ in Equation~\eqref{eq:likelihood} attains its maximum at  
\begin{equation} 
\hat{\bm{\beta}}_{\text{MLE}} = \argmax_{\bm{\beta}} \left\{ \sum^{T}_{t=p+1} \frac{1}{2} \left| y_t - \bm{y}^\prime_{t-1} \bm{\beta} \right|\right\}^{-(T-p)} = \argmin_{\bm{\beta}} \sum^{T}_{t=p+1} \frac{1}{2} \left| y_t - \bm{y}^\prime_{t-1} \bm{\beta} \right|,
\end{equation} 
provided $T > p$. This corresponds to the estimators of minimizing absolute error in median regression, which can be efficiently solved by linear programming~\citep{Koenker2005}. An analytical solution of $\hat{\tau}_{\text{MLE}}$ is available through a Gamma kernel:
\begin{align}
	\hat{\tau}_{\text{MLE}} & =\argmax_{\tau} \tau^{-(T-p)} \exp\left\{-\tau^{-1} \frac{1}{2} \sum^{T}_{t=p+1} \left| y_t - \bm{y}_{t-1}^\prime \hat{\bm{\beta}}_{\text{MLE}} \right| \right\} \\ 
	& = \frac{\frac{1}{2}\sum^T_{t=p+1} \left|y_t - \bm{y}^\prime_{t-1} \hat{\bm{\beta}}_{\text{MLE}} \right| }{T-p+1}.
\end{align} 
Before substituting $\hat{\bm{\beta}}_\text{MLE}$ and $\hat{\tau}_{\text{MLE}}$ into Equation~\eqref{eq:BIC.p}, we notice that both the likelihood function and sample size depend on the order $p$. To reconcile various sample sizes at different orders, we use the last $T-K$ samples to evaluate the likelihood function for any $p$. Consequently, $\text{BIC}_p$ is given by
\begin{equation}
	\text{BIC}_p = (p+2)\log(T-K) - 2\log\left[ (4\hat{\tau}_{\text{MLE}})^{-(T-K)} \exp \left( -\frac{1}{2\hat{\tau}_{\text{MLE}}} \sum^T_{t=K+1} | y_t -  \bm{y'}_{t-1} \hat{\bm{\beta}}_{\text{MLE}} |  \right) \right]. 
\end{equation}
The same methods to approach the unknown order and provide predictive densities apply to AR, where the likelihood function changes to Gaussian and the MLE of $\bm{\beta}$ that resembles the least square estimates has a simple closed-form expression. 

\section{Simulation}
\label{sec:simulation} 
In this section, we conduct simulations to assess the performances of BayesMAR with mean-based methods, focusing on parameter estimation under various model assumptions. To this end, we choose the AR model and the generalized autoregressive conditional heteroscedasticity model (GARCH), and defer predictive comparisons and more recent mean-based methods to real data application in Section~\ref{sec:Real}.

We generate data according to the model
\begin{equation}
y_t = \beta_0 + \beta_1 y_{t-1} + \beta_2 y_{t - 2} + \varepsilon_t, 
\end{equation}
where $\bm{\beta} = (\beta_0, \beta_1, \beta_2) = (0.3, 0.75, -0.35).$
We consider two scenarios depending on the distribution of $\varepsilon_t$: Gaussian error where $\varepsilon_t \sim N(0,1)$ and Laplace error where $\varepsilon_t \sim \Laplace(0,1)$, which correspond to the model assumptions of the AR and BayesMAR models, respectively. 

For each error assumption, we generate 200 observations and replicate such simulation 100 times. We estimate AR models via a Bayesian procedure with priors $\pi(\bm{\beta}) \propto 1$ and $\pi(\sigma) \propto \sigma^{-1}$. We also conduct the maximum likelihood estimation~\citep{GHP1980} for AR using the `\textit{arima}' function in the R package \texttt{stats}, which leads to almost identical performance and is thus not reported here. As such, we use AR and BayesAR exchangeably throughout this paper. We use AR($p$)-GARCH(1,1) when implementing GARCH, i.e., 
\begin{align}
	y_t & =  \beta_0 + \beta_1 y_{t-1} + \beta_2 y_{t-2} + \ldots + \beta_p y_{t-p} + e_t; \\
	e_t & =  \sqrt{h_t} \eta_t, \quad \eta_t \sim N(0,1); \\
	h_t & =  \omega_0 + \alpha_1 e^2_{t-1} + \alpha_2 h_{t-1}.
\end{align}
We fit the model using the R package \texttt{fGarch}, where all parameters are estimated by quasi-maximum likelihood~\citep{bollerslev1992quasi}. In addition, we implement the Quantile Autoregression (QAR) method proposed by~\cite{KX2006} using the R package \texttt{quantreg}, to compare its finite sample performance with BayesMAR. 

We assess estimates of $(\beta_0, \beta_1, \beta_2)$ by each method based on mean squared error (MSE). For a generic parameter $\theta$, letting the estimate be $\hat{\theta}_i$ in the $i$th simulation and $\bar{\theta} = \frac{1}{100} \sum^{100}_{i=1} \hat{{\theta}}_i$, then the MSE and its standard error are 
$$ \text{MSE} = \frac{1}{100} \sum_{i = 1}^{100} (\hat{\theta}_i - \theta)^2, \enskip \text{SE}_{\text{MSE}}  = \frac{1}{10}\sqrt{  \text{sample variance of } \{(\hat{\theta}_{i} - {\theta})^2\}_{i = 1}^{100}}.$$

We first provide the true order $p = 2$ to all models and compare their performances. Table~\ref{tab:MSE.simulation} reports the MSE of all methods. We can see that all methods benefit from a correctly specified error distribution: AR and GARCH have the smallest MSEs under Gaussian error, while BayesMAR and QAR have smaller MSEs when data are generated from Laplace distributions. However, BayesMAR appears to suffer less than AR from model misspecification; for example, the increase of MSE of $\beta_2$ under Gaussian error from AR to BayesMAR is 0.13, which is within two standard errors, while AR doubles the MSE of $\beta_2$, and so is beyond three standard errors of BayesMAR under Laplace error. It is reassuring that BayesMAR gives either the same or better MSEs than QAR in all cases, although all differences are within one standard error. This finite sample performance is consistent with the findings in~\cite{GCC2011} when comparing sampling-based Bayesian approaches with optimization-based counterparts for extreme quantile levels. 

\begin{table}
\captionsetup{width = 0.7\textwidth}
\caption{MSE of all methods under Gaussian error and Laplace error. Standard errors are reported below MSEs. All summaries have been multiplied by $10^2$. }
\label{tab:MSE.simulation}
\centering
\begin{tabular}{ccccccccccc}
 \Xhline{2\arrayrulewidth}
 \multicolumn{1}{c}{Models} & \multicolumn{2}{c}{Error} &  & \multicolumn{3}{c}{Gaussian} & & \multicolumn{3}{c}{Laplace} \\
 \hline
 \hline
 \multicolumn{1}{c}{} & \multicolumn{2}{c}{} & & $\beta_0$ & $\beta_1$ & $\beta_2$  & & $\beta_0$ & $\beta_1$ & $\beta_2$  \\
 \cline{5-7} \cline{9-11} 
    \multicolumn{1}{c}{\multirow{2}{*}{\shortstack{BayesMAR}}}& \multicolumn{2}{c}{MSE}&&   1.07 &  0.56 &  0.63 &&   0.73 &  0.27 &  0.19  \\ 
    \multicolumn{1}{c}{}&  \multicolumn{2}{c}{SE} &&    0.13 &  0.07 &  0.07  &&  0.11 &  0.04 &  0.03  \\ 
    \multicolumn{1}{c}{\multirow{2}{*}{\shortstack{QAR}}}&   \multicolumn{2}{c}{MSE} &&   1.21 &  0.62 &  0.75 &&   0.75 &  0.27 &  0.19  \\ 
    \multicolumn{1}{c}{}&  \multicolumn{2}{c}{SE} &&   0.15 &  0.08 &  0.08   &&   0.11 &  0.05 &  0.02\\ 
    \multicolumn{1}{c}{\multirow{2}{*}{\shortstack{AR}}} &
    \multicolumn{2}{c}{MSE}&&   0.77&  0.41&  0.50 &&   1.08 &  0.50& 0.43    \\ 
    \multicolumn{1}{c}{}&  \multicolumn{2}{c}{SE} &&   0.09 &  0.05 & 0.08    &&    0.16&   0.06&  0.06  \\ 
    \multicolumn{1}{c}{\multirow{2}{*}{\shortstack{GARCH}}}&  \multicolumn{2}{c}{MSE}&  &  0.76 &  0.41 &  0.51  &&   1.08 &  0.50 &  0.40  \\ 
    \multicolumn{1}{c}{}&  \multicolumn{2}{c}{SE} &&   0.09 &  0.05 &  0.08   &&   0.16 &  0.06 &  0.05  \\ 
 \Xhline{2\arrayrulewidth} 
\end{tabular}
\end{table} 

\begin{figure}[H]
\centering
\includegraphics[width=1\linewidth]{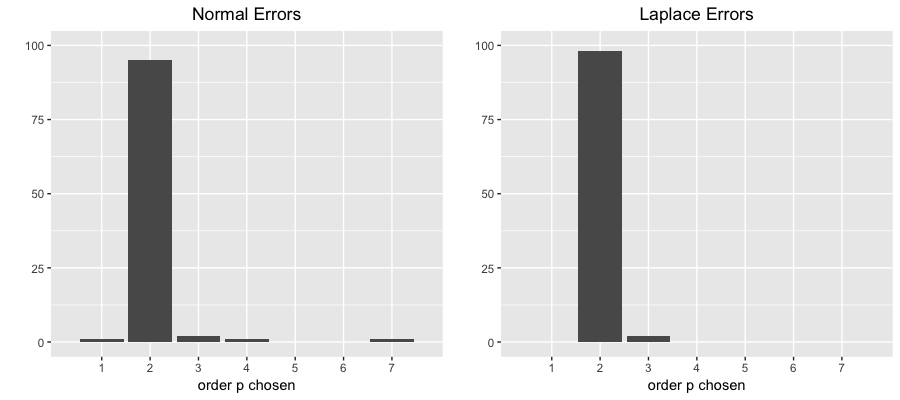}
\caption{Distributions of selected orders using BIC.}
\label{fig:sim_os}
\end{figure}

We next investigate the selection of the unknown order $p$ in BayesMAR using the BIC approach described in Section~\ref{sec:BIC}. We provide a large upper bound $K = 20$ for the order $p$. Figure~\ref{fig:sim_os} plots the distribution of $p$ in both scenarios and suggest that the selected orders almost always concentrate around the oracle value $p = 2$, even when the model is misspecified under Gaussian error. The overall accuracy across all simulations to select $p = 2$ using MAP is $95\%$ for normal errors and $98\%$ for Laplace errors.   

\section{Real Data Application}
\label{sec:Real}
\begin{figure}[H]
\centering
\includegraphics[width=0.8\linewidth]{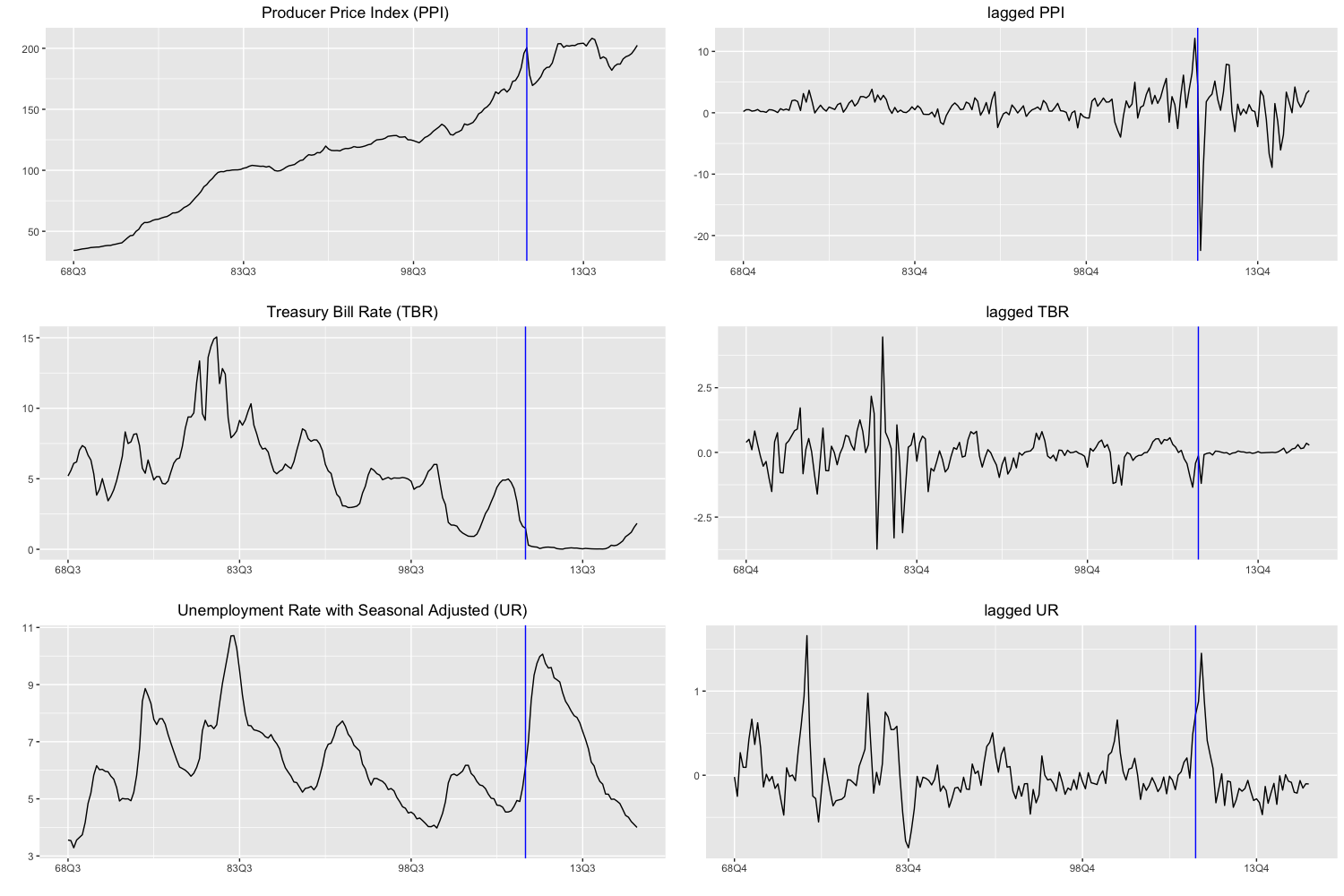}
\caption{Three time series (PPI, TBR, and UR) and the lagged data. The blue line is $t_0 = 08Q3$.}
\label{fig:data}
\end{figure}

In this section, we compare the predictive performances of the proposed methods to that of mean-based methods using various real-world data. We use three economic series from Federal Reserve Economic Data: the quarterly data Producer Price Index for all commodities~\citep{FRED:p}, 3-Month Treasury Bill: secondary market Rate~\citep{FRED:r}, and Unemployment Rate~\citep{FRED:u}, coded as PPI, TBR, and UR, respectively. Each time, the series ranges from 1968Q3 to 2018Q2, containing 200 observations. The unemployment rate data is seasonally adjusted by the method of seasonal-trend decomposition using Loess~\citep{CCMT1990}. The three time series $(y_t)$ and the lagged data of order one $(y_t - y_{t-1})$ are plotted in Figure~\ref{fig:data}. These three data sets have distinct patterns: the lagged PPI tends to be more stable before the crisis in the year of 2008, in contrast to the substantial fluctuation since then; the lagged TBR has more dramatic changes in earlier periods than in later periods; and the lagged UR appears to contain several extreme values, while a periodic pattern may still persist even after seasonal adjustment. These complex characteristics of the data enable a comparison between model-based methods when there is no guarantee for model assumptions to hold.

In addition to AR and GARCH, we implement selected methods proposed in research on the dynamic linear model. In particular, we consider the time-varying vector AR (TV-VAR) model proposed by~\cite{NW2013}, which links time-varying parameters to latent threshold processes and achieves state-of-the-art predictive performance in selected applications. A vector with a length of three that stacks the three time series $\bm{y}_t^{(3)}$, expands the dynamic linear model by utilizing a latent threshold vector $\bm{d} = (d_1, ..., d_k)$ with $d_i \geq 0$ for $i = 1,\ldots, k $ : 
\begin{align}
	 \bm{y}_t^{(3)} &= \bm{c}_t + \bm{B}_{1t}\bm{y}_{t-1}^{(3)} + \cdots + \bm{B}_{pt} \bm{y}_{t-p}^{(3)} +\bm{u}_t,\enskip \bm{u}_t \sim N(\bm{0}, \bm{\Sigma}_t),\enskip \bm{b}_t = (b_{1t},..., b_{kt})', \\
	b_{it} & =  \beta_{it} s_{it} \text{ with } s_{it} = \mathbbm{1}(|\beta_{it}| \ge d_i),  \\
	\bm{\beta}_{t} & =  \bm{\mu}_{\bm{\beta}} + \bm{\Phi}_{\bm{\beta}}(\bm{\beta}_{t-1} - \bm{\mu}_{\bm{\beta}}) + \bm{\eta}_{\bm{\beta}_t}, \enskip \bm{\eta}_{\bm{\beta}_t} \sim N(\bm{0}, \bm{V}_{\bm{\beta}}),
\end{align}
where $\bm{c}_t$ is the $3 \times 1$ vector of time-varying intercepts, $\bm{B}_{jt}$ is the $3 \times 3$ matrix of time-varying coefficients at lag $j$, $\bm{b}_t$ stacks $\bm{c}_t$ and $\bm{B}_{jt}$ by rows and by order $j$ from 1 to $p$, $k = 3(1 + 3p)$, and $\bm{\beta}_{t} = (\beta_{1t}, ..., \beta_{kt})$ is a latent time-varying parameter vector whose dynamic updates are specified by $(\bm{\mu}_{\bm{\beta}}, \bm{\Phi}_{\bm{\beta}}, \bm{V}_{\bm{\beta}})$ via a standard VAR model. The latent threshold vector $\bm{d}$ shrinks the time-varying coefficients $\bm{\beta}_t$ to zero, leading to dynamic sparsity and improved prediction. We implement two versions of TV-VAR: TV-VAR without latent threshold (NT) and TV-VAR with latent threshold (LT). Both methods are applicable to multivariate time series, and we adapt the OxMetrics code provided in \cite{NW2013} to the vector observations consisting of the three time series.  

For BayesMAR, we use \BMA~to denote the Bayesian model averaging strategy and \MAP~when the MAP estimate of $p$ is used, and we adopt the same convention for AR, i.e., BayesAR-BMA and BayesAR-MAP, which are implemented similarly to BayesMAR but with the Gaussian likelihood. 
GARCH(1,1) uses the same autoregressive order $p$ as chosen in BayesAR-MAP; although orders other than (1, 1) can be used, we observe that the GARCH(1,1) consistently led to the best predictive performances of GARCH in our real data application.
Since there are no immediately available model selection methods for NT or LT, we run the order from 1 to 5, and then choose the optimal results to favor NT and LT under each criterion. We observe that LT considerably outperformed NT in all scenarios, and thus only present the results for LT. 

We use recursive out-of-sample forecasting to assess each method after $t_0$=2008Q3. In particular, we fit each model with data up to quarter $t$ and conduct an $h$-step-ahead prediction for $h = 1, 2, 3, 4$. Then we move one period ahead and repeat the same procedure until we reach $t = T$. All methods are applied to the lagged data of order one to remove local trends and forecast the changes that lead to predictions of $y_t$. 

We calculate the root MSE (RMSE) and mean absolute error (MAE) from $t_0$ to $T$ to compare performance of each method, i.e., 
\begin{equation}
\text{RMSE} =  \sqrt{\frac{1}{35} \sum^{35}_{i=1} (y_{t_0+i} - \hat{y}_{t_0+i})^2}, \quad \text{MAE} = \frac{1}{35} \sum^{35}_{i=1} |y_{t_0+i} - \hat{y}_{t_0+i}|. 
\end{equation}
We additionally calculate the the relative change using \BMA~as the reference to ease comparison, that is, 
\begin{equation}
\text{Relative change} = \left[ \frac{\text{RMSE (or MAE) of each method }}{\text{RMSE (or MAE) of \BMA}} -1 \right] \times 100\%. 
\end{equation}
Table~\ref{tab:RMSE.Application} reports the RMSEs of all methods for the three data sets. The results indicate that BayesMAR outperforms alternatives in nearly all scenarios. In particular, the RMSEs of \BMA~is uniformly smaller than its AR counterpart. Compared to \BMA, the methods of AR, GARCH, and LT increase the RMSE by 10\% to 89.2\% for TBR, and 3.0\% to 35.6\% for UR, with two exceptions for the UR when compared to GARCH at three- and four-steps-ahead predictions where GARCH slightly reduces the RMSE by 0.3\% and 2.8\%, respectively. Although the gap in the prediction for PPI is smaller (between 0.8\% to 8.9\%), all competing methods have a larger RMSE than \BMA. GARCH accounts for comprehensive variance structures and LT dynamically updates regression coefficients, which are arguably powerful methods with considerable complexity; it is remarkable that the proposed simple BayesMAR leads to favorable and often superior performance using real-world data. 

\begin{table}
\captionsetup{width = 0.85\textwidth}
\caption{Forecasting Performance for U.S. macroeconomic data after 2007-2008 financial crisis: RMSEs for $h$-step ahead prediction for $h = 1, 2, 3, 4$. The relative change uses \BMA~as the baseline. RMSEs for Treasury Bill Rate and Unemployment Rate have been multiplied by $10$.}
\centering
{  \begin{tabular}{cccccccccccc}
 \Xhline{2\arrayrulewidth}
\multicolumn{12}{c}{Producer Price Index} \\
\hline
\hline
\multicolumn{2}{c}{ \multirow{2}{*}{Models}}  & &
\multicolumn{4}{c}{ RMSE } &  & \multicolumn{4}{c}{ Relative change (\%)   }  \\ 
\cline{4-7} \cline{9-12}
 \multicolumn{2}{c}{} & & 1 & 2 & 3 & 4 && 1 & 2 & 3 & 4\\
\cline{1-2}\cline{4-7} \cline{9-12}
  \multicolumn{2}{c}{\BMA}&& 3.18 & 5.94 & 7.74 & 9.24 &&  - & - &  - &  - \\
  \multicolumn{2}{c}{\MAP}&& 3.21 & 5.94 & 7.69 & 9.13 && 0.7 & 0.1 & -0.7 & -1.1 \\ 
  \multicolumn{2}{c}{\arBMA}&& 3.37 & 6.24 & 8.04 & 9.39 &&  6.0 & 5.1 & 3.8 & 1.6 \\ 
  \multicolumn{2}{c}{\arMAP}&& 3.38 & 6.26 & 8.07 & 9.43 && 6.0 & 5.4 & 4.2 & 2.1 \\ 
  \multicolumn{2}{c}{GARCH}&& 3.21 & 6.05 & 7.85 & 9.31 &&  0.8 & 1.9 & 1.4 & 0.8 \\ 
  \multicolumn{2}{c}{LT}&& 3.47 & 6.06 & 7.94 & 9.45   && 8.9 & 2.0 & 2.5 & 2.3 \\ 
 \Xhline{2\arrayrulewidth}
\multicolumn{12}{c}{Treasury Bill Rate} \\
\hline
\hline
\multicolumn{2}{c}{ \multirow{2}{*}{Models}}  & &
\multicolumn{4}{c}{ RMSE } &  & \multicolumn{4}{c}{ Relative change (\%)   }  \\ 
\cline{4-7} \cline{9-12}
 \multicolumn{2}{c}{} & & 1 & 2 & 3 & 4 && 1 & 2 & 3 & 4\\
\cline{1-2}\cline{4-7} \cline{9-12}
  \multicolumn{2}{c}{\BMA}&& 0.91 & 1.49 & 2.16 & 2.94  &&  - & - &  - &  - \\
  \multicolumn{2}{c}{\MAP}&& 0.90 & 1.48 & 2.16 & 2.93 &&  -0.9 & -0.1 & -0.1 & -0.1  \\ 
  \multicolumn{2}{c}{\arBMA}&& 1.63 & 2.10 & 2.99 & 4.16  && 79.7 & 41.6 & 38.2 & 41.7 \\ 
  \multicolumn{2}{c}{\arMAP}&& 1.71 & 2.37 & 3.31 & 4.56 && 89.2 & 59.3 & 53.1 & 55.1 \\ 
  \multicolumn{2}{c}{GARCH}&& 1.66 & 2.02 & 2.63 & 3.4 &&  83.6 & 35.9 & 21.7 & 18.5 \\ 
 \multicolumn{2}{c}{LT}&& 1.00 & 1.87 & 2.66 & 3.55 && 10.3 & 25.6 & 23.0 & 21.0\\ 
 \Xhline{2\arrayrulewidth}
\multicolumn{12}{c}{Unemployment Rate } \\
\hline
\hline
\multicolumn{2}{c}{ \multirow{2}{*}{Models}}  & &
\multicolumn{4}{c}{ RMSE } &  & \multicolumn{4}{c}{ Relative change (\%)   }  \\ 
\cline{4-7} \cline{9-12}
 \multicolumn{2}{c}{} & & 1 & 2 & 3 & 4 && 1 & 2 & 3 & 4\\
\cline{1-2}\cline{4-7} \cline{9-12}
  \multicolumn{2}{c}{\BMA}&&  1.76 & 3.07 & 4.34 & 5.85&&  - & - &  - &  - \\
  \multicolumn{2}{c}{\MAP}&&  1.77 & 3.11 & 4.42 & 5.94  && 0.5 & 1.2 & 1.8 & 1.6   \\ 
  \multicolumn{2}{c}{\arBMA}&& 1.88 & 3.37 & 4.88 & 6.88 && 7.2 & 9.5 & 12.3 & 17.6 \\ 
  \multicolumn{2}{c}{\arMAP}&&  1.88 & 3.42 & 5.11 & 7.18 && 6.8 & 11.3 & 17.5 & 22.8 \\ 
  \multicolumn{2}{c}{GARCH}&& 1.86 & 3.17 & 4.33 & 5.69  &&  5.6 & 3.0 & -0.3 & -2.8  \\
  \multicolumn{2}{c}{LT}&& 2.40 & 4.17 & 5.53 & 7.17   &&36.7 & 35.6 & 27.4 & 22.7 \\
 \Xhline{2\arrayrulewidth}
\end{tabular}}
 \label{tab:RMSE.Application}
\end{table}

Table~\ref{tab:MAE.Application} reports the MAEs of all methods, suggesting similar observations as those made from Table~\ref{tab:RMSE.Application}. The proposed BayesMAR methods give the smallest MAEs in nearly all scenarios, with one exception for the UR when compared to GARCH. 

\begin{table}
\captionsetup{width = 0.85\textwidth}
\caption{Forecasting Performance for U.S. macroeconomic data after 2007-2008 financial crisis: MAEs for $h$-step ahead prediction for $h = 1, 2, 3, 4$. The relative change uses \BMA~as the baseline.  MAEs for Treasury Bill Rate and Unemployment Rate have been multiplied by $10$.}
\centering
{\begin{tabular}{cccccccccccc}
 \Xhline{2\arrayrulewidth}
\multicolumn{12}{c}{Producer Price Index} \\
\hline
\hline
\multicolumn{2}{c}{ \multirow{2}{*}{Models}}  & &
\multicolumn{4}{c}{ MAE } &  & \multicolumn{4}{c}{ Relative change (\%)   }  \\ 
\cline{4-7} \cline{9-12}
 \multicolumn{2}{c}{} & & 1 & 2 & 3 & 4 && 1 & 2 & 3 & 4\\
\cline{1-2}\cline{4-7} \cline{9-12}
  \multicolumn{2}{c}{\BMA}&& 2.50 & 4.44 & 5.90 & 7.26  &&  - & - &  - &  - \\
  \multicolumn{2}{c}{\MAP}&&2.50 & 4.44 & 5.84 & 7.18  &&  0.2 & 0.1 & -1.0 & -1.1  \\ 
  \multicolumn{2}{c}{\arBMA}&& 2.69 & 4.49 & 6.05 & 7.33  && 7.8 & 1.2 & 2.5 & 1.0 \\ 
  \multicolumn{2}{c}{\arMAP}&& 2.70 & 4.54 & 6.05 & 7.37 && 8.0 & 2.2 & 2.5 & 1.6\\ 
  \multicolumn{2}{c}{GARCH}&& 2.54 & 4.59 & 5.98 & 7.29 &&  1.8 & 3.5 & 1.4 & 0.4  \\ 
  \multicolumn{2}{c}{LT}&& 2.72 & 4.47 & 6.29 & 7.45    && 8.9 & 0.8 & 6.6 & 2.6 \\ 
\Xhline{2\arrayrulewidth}
\multicolumn{12}{c}{Treasury Bill Rate} \\
\hline
\hline
\multicolumn{2}{c}{ \multirow{2}{*}{Models}}  & &
\multicolumn{4}{c}{ MAE } &  & \multicolumn{4}{c}{ Relative change (\%)   }  \\ 
\cline{4-7} \cline{9-12}
 \multicolumn{2}{c}{} & & 1 & 2 & 3 & 4 && 1 & 2 & 3 & 4\\
\cline{1-2}\cline{4-7} \cline{9-12}
  \multicolumn{2}{c}{\BMA}&& 0.56 & 1.05 & 1.56 & 2.10   &&  - & - &  - &  - \\
  \multicolumn{2}{c}{\MAP}&&  0.56 & 1.05 & 1.56 & 2.10  &&  -0.6 & -0.1 &  0.0 & 0.0  \\ 
  \multicolumn{2}{c}{\arBMA}&& 1.09 & 1.61 & 2.25 & 3.04  &&  95.5 & 52.7 & 44.2 & 44.6 \\ 
  \multicolumn{2}{c}{\arMAP}&& 1.14 & 1.80 & 2.52 & 3.43  && 103.2 & 70.5 & 61.6 & 63.1  \\ 
  \multicolumn{2}{c}{GARCH}&& 0.99 & 1.43 & 2.01 & 2.68  &&  77.8 & 35.7 & 28.7 & 27.6 \\ 
  \multicolumn{2}{c}{LT}&&  0.65 & 1.23 & 1.78 & 2.37  && 16.0 & 16.2 & 13.9 & 13.0 \\ 
\Xhline{2\arrayrulewidth}
\multicolumn{12}{c}{Unemployment Rate } \\
\hline
\hline
\multicolumn{2}{c}{ \multirow{2}{*}{Models}}  & &
\multicolumn{4}{c}{ MAE } &  & \multicolumn{4}{c}{ Relative change (\%)   }  \\ 
\cline{4-7} \cline{9-12}
 \multicolumn{2}{c}{} & & 1 & 2 & 3 & 4 && 1 & 2 & 3 & 4\\
\cline{1-2}\cline{4-7} \cline{9-12}
  \multicolumn{2}{c}{\BMA}&&  1.24 & 2.30 & 3.49 & 4.85 &&  - & - &  - &  - \\
  \multicolumn{2}{c}{\MAP}&& 1.26 & 2.34 & 3.57 & 4.95 && 1.2 & 1.4 & 2.3 & 2.1   \\ 
  \multicolumn{2}{c}{\arBMA}&&  1.47 & 2.61 & 3.91 & 5.73  &&18.1 & 13.1 & 12.2 & 18.2\\ 
  \multicolumn{2}{c}{\arMAP}&& 1.49 & 2.69 & 4.16 & 5.94  &&19.8 & 16.9 & 19.2 & 22.4  \\ 
  \multicolumn{2}{c}{GARCH}&& 1.27 & 2.12 & 3.16 & 4.27  && 2.4 & -8.1 & -9.5 & -11.9  \\ 
  \multicolumn{2}{c}{LT}&& 1.66 & 3.26 & 4.63 & 6.32  && 33.4 & 41.7 & 32.6 & 30.4 \\  
 \hline
 \hline
\end{tabular}
\label{tab:MAE.Application}}
\end{table}

The RMSEs of \BMA~and \MAP~are close to each other with at most 1.1\% (PPI), -0.6\% (TBR), 2.3\% (UR) relative differences, and similar observations hold for MAEs. This is partly caused by highly concentrated weights of model orders when using BIC for order selection. In particular, we find \BMA~puts most weight on the order selected in \MAP, leading to minimal differences between the two variants of BayesMAR.  Comparing BayesAR-BMA and BayesAR-MAP leads to the same conclusion for AR. 

\begin{figure}
\centering
\includegraphics[width=0.95\linewidth]{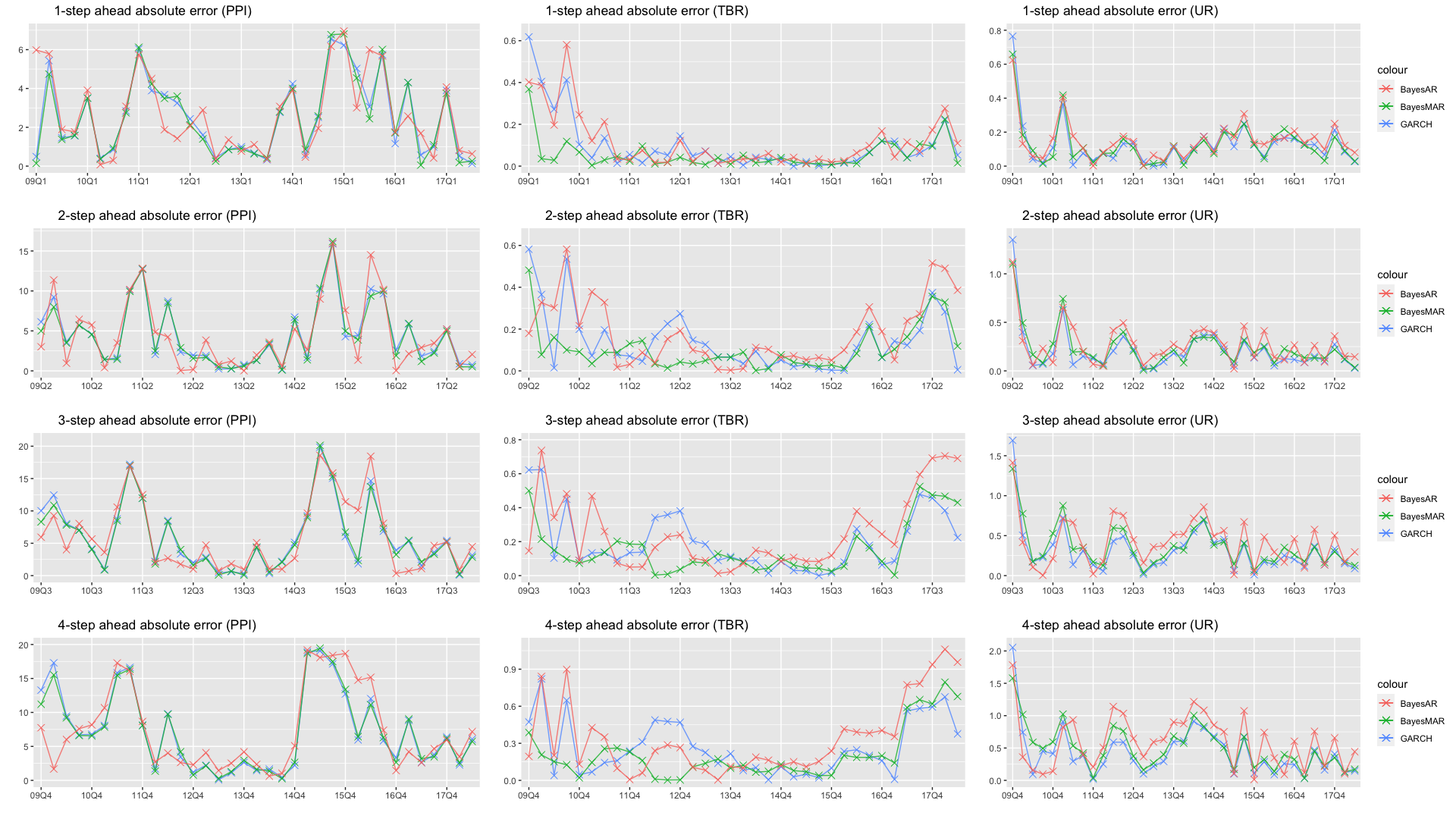}
\caption{   Absolute $h$-step ahead predictive errors at each period from 2008Q3 for $h = 1, 2, 3, 4$. Orders of BayesMAR and BayesAR are selected by the BIC.}
\label{fig:cp_1saf}
\end{figure}

Figure~\ref{fig:cp_1saf} compares up to four-steps-ahead absolute predictive errors at each $t$ from $t_0$ to $T$ given by the three static models, MAR, AR, and GARCH, which provides insights into understanding the performance of BayesMAR. We choose AR and GARCH and implement MAR and AR using MAP such that all methods in the figure build on similar model structures. We can see that AR yields a large prediction error at the beginning (PPI), which is substantially reduced by GARCH, which incorporates heterogeneous variance structures. It is reassuring that the proposed BayesMAR method, which uses a simple constant variance structure, achieves the same predictive gain (for PPI) or even further improvements (for TBR and UR). At later time points when the time series stabilizes without usually large deviations, BayesMAR tends to perform similarly to AR and GARCH. Since BayesMAR is a parametric model bearing the same model structure as AR, these comparisons suggest that further performance gains may be possible by following the rich literature that extends AR to more advanced models such as GARCH and dynamic models, by simply altering the error assumption from Gaussian to Laplace. 

\begin{figure}
\centering
\includegraphics[width=1\linewidth]{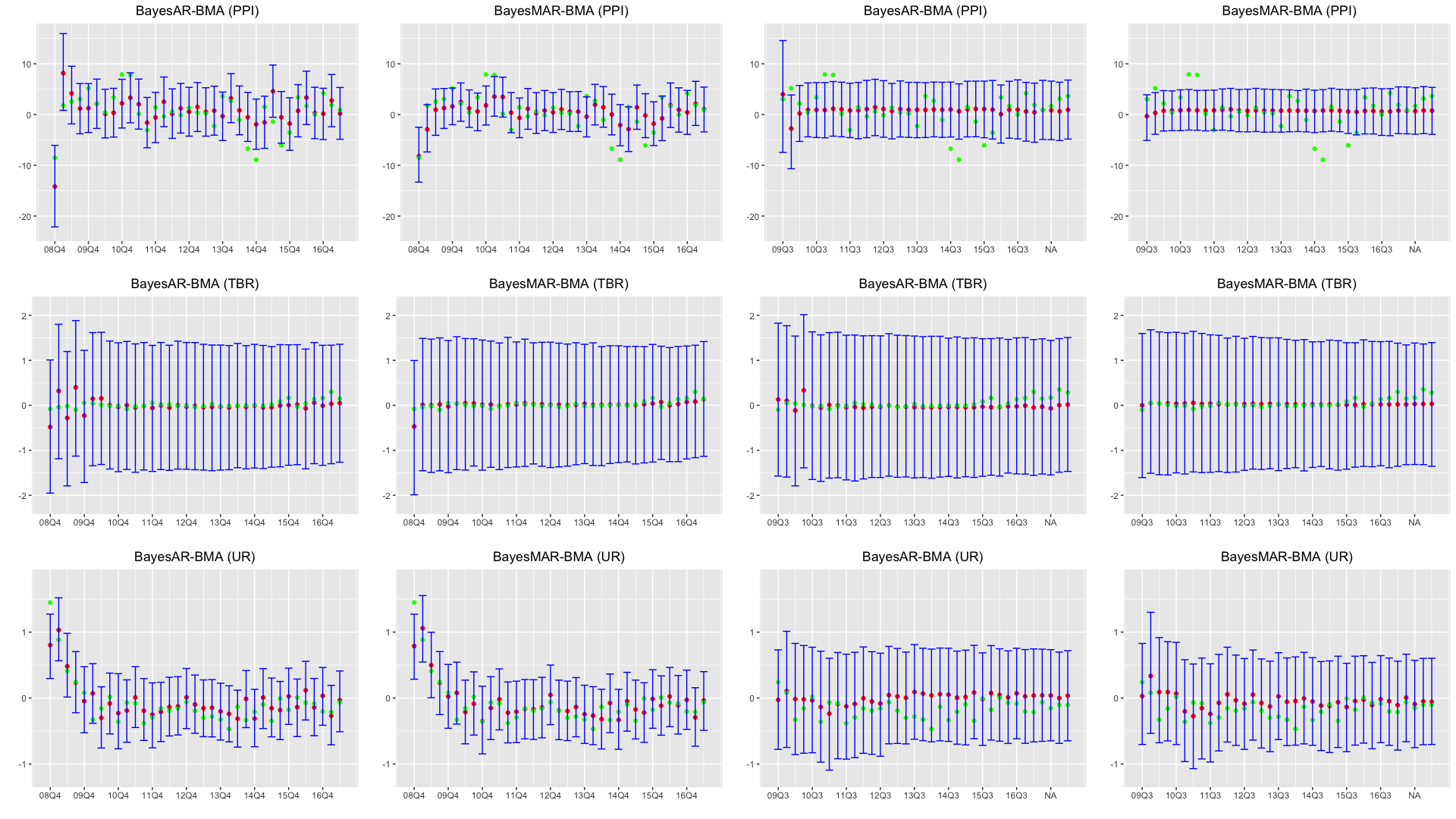}
\caption{1-step ahead (the first two columns) and 4-step ahead (the last two columns) 95\% credible intervals of MAR and AR using BMA. True observations are points marked green, and predictions are marked red. Blue bars are the $95\%$ credible intervals.} 
\label{fig:CIplots}
\end{figure}

We plot the 95\% credible intervals of AR and MAR in Figure~\ref{fig:CIplots}, both using the BMA version. For both AR and MAR, four-steps-ahead predictive intervals appear wider than one-step-ahead predictive intervals. This makes sense, as more uncertainty propagates, and this is particularly the case for the UR data. MAR is more robust than AR at the early stage of the PPI time series. Since an informative comparison between MAR and AR through this visualization seems difficult, we next turn to numerical comparisons. 

\color{black} 

The proposed method provides density forecasts in addition to point forecasts; see Section~\ref{sec:method}. We evaluate the probabilistic forecasts produced by the proposed method and AR through the continuous ranked probability score (CRPS)~\citep{TFA2007}, which is implemented in the R package \texttt{scoringRules}. We can see that the proposed MAR model produces a considerably smaller CRPS than AR for the TBR and UR data, while being close to AR for PPI. This suggests that the advantage of MAR may extend to probabilistic forecasts beyond point forecasts (see Table~\ref{tab:CRPS}). 

   \begin{table}[H]
	\setlength{\extrarowheight}{3pt}
	\captionsetup{width = 0.85\textwidth}
	\caption{Forecasting Performance for U.S. macroeconomic data after 2007-2008 financial crisis: CRPS for $h$-step ahead prediction for $h = 1, 2, 3, 4$. The relative change uses \BMA~as the baseline. CRPS for Treasury Bill Rate and Unemployment Rate have been multiplied by $10$.}
	\centering
	{ \begin{tabular}{cccccccccccc}
			\Xhline{2\arrayrulewidth}
			\multicolumn{12}{c}{Producer Price Index} \\
			\hline
			\hline
			\multicolumn{2}{c}{ \multirow{2}{*}{Models}}  & &
			\multicolumn{4}{c}{ CRPS } &  & \multicolumn{4}{c}{ Relative change (\%)   }  \\ 
			\cline{4-7} \cline{9-12}
			\multicolumn{2}{c}{} & & 1 & 2 & 3 & 4 && 1 & 2 & 3 & 4\\
			\cline{1-2}\cline{4-7} \cline{9-12}
			\multicolumn{2}{c}{\BMA } &&1.93 & 2.12 & 2.02 & 2.01  &&  - & - & - & - \\ 
			\multicolumn{2}{c}{\MAP}&&  1.95 & 2.13 & 2.00 & 2.00 && 1.1 & 0.8 & -0.8 & -0.8 \\ 
			\multicolumn{2}{c}{\arBMA}&& 1.93 & 2.08 & 2.13 & 2.05 && 0.1 & -1.8 & 5.7 & 2.1 \\
			\multicolumn{2}{c}{\arMAP}&& 1.93 & 2.08 & 2.14 & 2.06 && 0.3 & -1.5 & 6.1 & 2.6 \\
			\Xhline{2\arrayrulewidth}
			\multicolumn{12}{c}{Treasury Bill Rate} \\
			\hline
			\hline
			\multicolumn{2}{c}{ \multirow{2}{*}{Models}}  & &
			\multicolumn{4}{c}{ CRPS } &  & \multicolumn{4}{c}{ Relative change (\%)   }  \\ 
			\cline{4-7} \cline{9-12}
			\multicolumn{2}{c}{} & & 1 & 2 & 3 & 4 && 1 & 2 & 3 & 4\\
			\cline{1-2}\cline{4-7} \cline{9-12}
			\multicolumn{2}{c}{\BMA } && 1.22 & 1.37 & 1.43 & 1.45 &&  - & - & - & - \\ 
			\multicolumn{2}{c}{\MAP}&& 1.22 & 1.37 & 1.42 & 1.45 && 0.0 & 0.1 & -0.3 & 0.0 \\ 
			\multicolumn{2}{c}{\arBMA}&& 1.80 & 1.89 & 1.92 & 1.97 && 48.4 & 37.5 & 34.2 & 35.6 \\ 
			\multicolumn{2}{c}{\arMAP}&& 1.81 & 1.88 & 1.91 & 1.97 && 48.5 & 37.3 & 33.8 & 35.7  \\ 
			\Xhline{2\arrayrulewidth}
			\multicolumn{12}{c}{Unemployment Rate} \\
			\hline
			\hline
			\multicolumn{2}{c}{ \multirow{2}{*}{Models}}  & &
			\multicolumn{4}{c}{ CRPS } &  & \multicolumn{4}{c}{ Relative change (\%)   }  \\ 
			\cline{4-7} \cline{9-12}
			\multicolumn{2}{c}{} & & 1 & 2 & 3 & 4 && 1 & 2 & 3 & 4\\
			\cline{1-2}\cline{4-7} \cline{9-12}
			\multicolumn{2}{c}{\BMA } && 0.93 & 0.97 & 1.04 & 1.16 &&  - & - & - & - \\ 
			\multicolumn{2}{c}{\MAP}&& 0.93 & 0.98 & 1.05 & 1.16 && 0.7 & 0.7 & 0.7 & 0.5 \\ 
			\multicolumn{2}{c}{\arBMA}&& 1.04 & 1.14 & 1.22 & 1.42  && 12.6 & 17.0 & 17.9 & 22.2 \\ 
			\multicolumn{2}{c}{\arMAP}&& 1.04 & 1.13 & 1.26 & 1.45 && 11.9 & 16.5 & 21.1 & 24.7 \\ 
		    \Xhline{2\arrayrulewidth}
	\end{tabular}}
	\label{tab:CRPS}
\end{table}

 \color{black}

\section{Concluding remarks} \label{sec:conclusion}
This article proposes a Bayesian median autoregressive (BayesMAR) model for robust time series forecasting. The proposed method has close connections with time-varying quantile regression. BayesMAR adopts a parametric model bearing the same structure as AR models by altering the Gaussian error to Laplace, leading to a simple, robust, and interpretable modeling strategy for time series forecasting with principled uncertainty quantification through Bayesian model averaging and Bayesian model selection. Real data applications using U.S. macroeconomic data show that BayesMAR leads to favorable and often superior predictive performances compared to the selected state-of-the-art mean-based alternatives under various loss functions that encompass both point and probabilistic forecasts.

BayesMAR enjoys practical benefits as a technical tool to introduce robustness and improve predictions. In addition, since the autoregressive structure in BayesMAR resembles the widely used AR model, BayesMAR can be used to complement a rich class of methods that build on the AR model. The AR model is arguably one of the most popular methods in time series, serving as the building block for other models such as GARCH and TV-VAR in research on the dynamic linear model. The proposed MAR model shows the potential for further performance gains by following the rich literature that extends AR to more advanced models with Laplace error assumptions rather than Gaussian ones. 
	
\section*{Declaration of competing interest}
The authors declare that they have no known competing financial interests or personal relationships that could have appeared to influence the work reported in this paper.

\section*{Acknowledgements}
This work was partially supported by the grant DMS-2015569 from the National Science Foundation.

\section*{Data availability}
The R code to implement the proposed methods is publicly available at \url{https://github.com/xylimeng/BayesMAR}.

\bibliography{BayesCAViaR.bib}

\begin{thebibliography}{50}
\expandafter\ifx\csname natexlab\endcsname\relax\def\natexlab#1{#1}\fi
\providecommand{\url}[1]{\texttt{#1}}
\providecommand{\href}[2]{#2}
\providecommand{\path}[1]{#1}
\providecommand{\DOIprefix}{doi:}
\providecommand{\ArXivprefix}{arXiv:}
\providecommand{\URLprefix}{URL: }
\providecommand{\Pubmedprefix}{pmid:}
\providecommand{\doi}[1]{\href{http://dx.doi.org/#1}{\path{#1}}}
\providecommand{\Pubmed}[1]{\href{pmid:#1}{\path{#1}}}
\providecommand{\bibinfo}[2]{#2}
\ifx\xfnm\relax \def\xfnm[#1]{\unskip,\space#1}\fi
\bibitem[{{Akouemo} \& {Povinelli}(2014)}]{HR2014}
\bibinfo{author}{{Akouemo}, H.~N.}, \& \bibinfo{author}{{Povinelli}, R.~J.}
  (\bibinfo{year}{2014}).
\newblock \bibinfo{title}{Time series outlier detection and imputation}.
\newblock In {\it \bibinfo{booktitle}{Proceedings of IEEE PES General Meeting
  \textbar Conference \& Exposition}\/} (pp. \bibinfo{pages}{1--5}).
\bibitem[{Bollerslev(1986)}]{bollerslev1986}
\bibinfo{author}{Bollerslev, T.} (\bibinfo{year}{1986}).
\newblock \bibinfo{title}{Generalized autoregressive conditional
  heteroskedasticity}.
\newblock {\it \bibinfo{journal}{Journal of Econometrics}\/},  {\it
  \bibinfo{volume}{31}\/}, \bibinfo{pages}{307--327}.
\bibitem[{Bollerslev \& Wooldridge(1992)}]{bollerslev1992quasi}
\bibinfo{author}{Bollerslev, T.}, \& \bibinfo{author}{Wooldridge, J.~M.}
  (\bibinfo{year}{1992}).
\newblock \bibinfo{title}{Quasi-maximum likelihood estimation and inference in
  dynamic models with time-varying covariances}.
\newblock {\it \bibinfo{journal}{Econometric Reviews}\/},  {\it
  \bibinfo{volume}{11}\/}, \bibinfo{pages}{143--172}.
\bibitem[{Cameron \& Pettitt(2014)}]{cp2014}
\bibinfo{author}{Cameron, E.}, \& \bibinfo{author}{Pettitt, A.}
  (\bibinfo{year}{2014}).
\newblock \bibinfo{title}{Recursive pathways to marginal likelihood estimation
  with prior-sensitivity analysis}.
\newblock {\it \bibinfo{journal}{Statistical Science}\/},  {\it
  \bibinfo{volume}{29}\/}, \bibinfo{pages}{397--419}.
\bibitem[{Casini \& Perron(2019)}]{CP2019}
\bibinfo{author}{Casini, A.}, \& \bibinfo{author}{Perron, P.}
  (\bibinfo{year}{2019}).
\newblock \bibinfo{title}{Structural breaks in time series}.
\newblock {\it \bibinfo{journal}{Oxford Research Encyclopedia of Economics and
  Finance}\/}, .
\bibitem[{Chen \& Liu(1993{\natexlab{a}})}]{CL1993}
\bibinfo{author}{Chen, C.}, \& \bibinfo{author}{Liu, L.-M.}
  (\bibinfo{year}{1993}{\natexlab{a}}).
\newblock \bibinfo{title}{Forecasting time series with outliers}.
\newblock {\it \bibinfo{journal}{Journal of Forecasting}\/},  {\it
  \bibinfo{volume}{12}\/}, \bibinfo{pages}{13--35}.
\bibitem[{Chen \& Liu(1993{\natexlab{b}})}]{CLJ1993}
\bibinfo{author}{Chen, C.}, \& \bibinfo{author}{Liu, L.-M.}
  (\bibinfo{year}{1993}{\natexlab{b}}).
\newblock \bibinfo{title}{Joint estimation of model parameters and outlier
  effects in time series}.
\newblock {\it \bibinfo{journal}{Journal of the American Statistical
  Association}\/},  {\it \bibinfo{volume}{88}\/}, \bibinfo{pages}{284--297}.
\bibitem[{Chen \& So(2006)}]{CS2006}
\bibinfo{author}{Chen, C.~W.}, \& \bibinfo{author}{So, M.~K.}
  (\bibinfo{year}{2006}).
\newblock \bibinfo{title}{On a threshold heteroscedastic model}.
\newblock {\it \bibinfo{journal}{International Journal of Forecasting}\/},
  {\it \bibinfo{volume}{22}\/}, \bibinfo{pages}{73--89}.
\bibitem[{Choi \& Hobert(2013)}]{CH2013}
\bibinfo{author}{Choi, H.~M.}, \& \bibinfo{author}{Hobert, J.~P.}
  (\bibinfo{year}{2013}).
\newblock \bibinfo{title}{Analysis of mcmc algorithms for bayesian linear
  regression with laplace errors}.
\newblock {\it \bibinfo{journal}{Journal of Multivariate Analysis}\/},  {\it
  \bibinfo{volume}{117}\/}, \bibinfo{pages}{32--40}.
\bibitem[{Cleveland et~al.(1990)Cleveland, Cleveland, McRae \&
  Terpenning}]{CCMT1990}
\bibinfo{author}{Cleveland, R.~B.}, \bibinfo{author}{Cleveland, W.~S.},
  \bibinfo{author}{McRae, J.~E.}, \& \bibinfo{author}{Terpenning, I.}
  (\bibinfo{year}{1990}).
\newblock \bibinfo{title}{Stl: A seasonal-trend decomposition}.
\newblock {\it \bibinfo{journal}{Journal of official statistics}\/},  {\it
  \bibinfo{volume}{6}\/}, \bibinfo{pages}{3--73}.
\bibitem[{Croux et~al.(2008)Croux, Gelper \& Fried}]{CGF2008}
\bibinfo{author}{Croux, C.}, \bibinfo{author}{Gelper, S.}, \&
  \bibinfo{author}{Fried, R.} (\bibinfo{year}{2008}).
\newblock \bibinfo{title}{Computational aspects of robust holt-winters
  smoothing based on m-estimation}.
\newblock {\it \bibinfo{journal}{Applications of Mathematics}\/},  {\it
  \bibinfo{volume}{53}\/}, \bibinfo{pages}{163}.
\bibitem[{Engle \& Manganelli(2004)}]{EM2004}
\bibinfo{author}{Engle, R.~F.}, \& \bibinfo{author}{Manganelli, S.}
  (\bibinfo{year}{2004}).
\newblock \bibinfo{title}{Caviar: Conditional autoregressive value at risk by
  regression quantiles}.
\newblock {\it \bibinfo{journal}{Journal of Business \& Economic
  Statistics}\/},  {\it \bibinfo{volume}{22}\/}, \bibinfo{pages}{367--381}.
\bibitem[{Fan \& Yao(2008)}]{fan2008nonlinear}
\bibinfo{author}{Fan, J.}, \& \bibinfo{author}{Yao, Q.} (\bibinfo{year}{2008}).
\newblock {\it \bibinfo{title}{Nonlinear Time Series: Nonparametric and
  Parametric Methods}\/}.
\newblock \bibinfo{publisher}{Springer Science \& Business Media}.
\bibitem[{Ferraty \& Vieu(2006)}]{ferraty2006nonparametric}
\bibinfo{author}{Ferraty, F.}, \& \bibinfo{author}{Vieu, P.}
  (\bibinfo{year}{2006}).
\newblock {\it \bibinfo{title}{Nonparametric Functional Data Analysis: Theory
  and Practice}\/}.
\newblock \bibinfo{publisher}{Springer Science \& Business Media}.
\bibitem[{Fox(1972)}]{Fox1972}
\bibinfo{author}{Fox, A.~J.} (\bibinfo{year}{1972}).
\newblock \bibinfo{title}{Outliers in time series}.
\newblock {\it \bibinfo{journal}{Journal of the Royal Statistical Society:
  Series B (Methodological)}\/},  {\it \bibinfo{volume}{34}\/},
  \bibinfo{pages}{350--363}.
\bibitem[{{FRED: Board of Governors of the Federal Reserve System
  (US)}(2018)}]{FRED:r}
\bibinfo{author}{{FRED: Board of Governors of the Federal Reserve System (US)}}
  (\bibinfo{year}{2018}).
\newblock \bibinfo{title}{{3-Month Treasury Bill: Secondary Market Rate}}.
\newblock \bibinfo{note}{Retrieved from FRED, Federal Reserve Bank of St.
  Louis; https://fred.stlouisfed.org/series/TB3MS, Aug 09, 2018.}
\bibitem[{{FRED: Organization for Economic Co-operation and
  Development}(2018)}]{FRED:u}
\bibinfo{author}{{FRED: Organization for Economic Co-operation and
  Development}} (\bibinfo{year}{2018}).
\newblock \bibinfo{title}{{Unemployment Rate: Aged 15-64: All Persons for the
  United States}}.
\newblock \bibinfo{note}{Retrieved from FRED, Federal Reserve Bank of St.
  Louis; https://fred.stlouisfed.org/series/LRUN64TTUSQ156N, Aug 09, 2018.}
\bibitem[{{FRED: U.S. Bureau of Labor Statistics}(2018)}]{FRED:p}
\bibinfo{author}{{FRED: U.S. Bureau of Labor Statistics}}
  (\bibinfo{year}{2018}).
\newblock \bibinfo{title}{{Producer Price Index for All Commodities}}.
\newblock \bibinfo{note}{Retrieved from FRED, Federal Reserve Bank of St.
  Louis; https://fred.stlouisfed.org/series/PPIACO, Aug 09, 2018.}
\bibitem[{Gardner et~al.(1980)Gardner, Harvey \& Phillips}]{GHP1980}
\bibinfo{author}{Gardner, G.}, \bibinfo{author}{Harvey, A.~C.}, \&
  \bibinfo{author}{Phillips, G. D.~A.} (\bibinfo{year}{1980}).
\newblock \bibinfo{title}{An algorithm for exact maximum likelihood estimation
  of autoregressive-moving average models by means of {K}alman filtering}.
\newblock {\it \bibinfo{journal}{Journal of the Royal Statistical Society.
  Series C (Applied Statistics)}\/},  {\it \bibinfo{volume}{29}\/},
  \bibinfo{pages}{311--322}.
\bibitem[{Gelman et~al.(1996)Gelman, Roberts \& Gilks}]{GRG1996}
\bibinfo{author}{Gelman, A.}, \bibinfo{author}{Roberts, G.~O.}, \&
  \bibinfo{author}{Gilks, W.~R.} (\bibinfo{year}{1996}).
\newblock \bibinfo{title}{Efficient metropolis jumping rules}.
\newblock In \bibinfo{editor}{J.~M. Bernardo}, \bibinfo{editor}{J.~O. Berger},
  \bibinfo{editor}{A.~P. Dawid}, \& \bibinfo{editor}{A.~F.~M. Smith} (Eds.),
  {\it \bibinfo{booktitle}{Bayesian Statistics 5}\/} (pp.
  \bibinfo{pages}{599--607}).
\newblock \bibinfo{publisher}{Oxford}.
\bibitem[{Gerlach et~al.(2011)Gerlach, Chen \& Chan}]{GCC2011}
\bibinfo{author}{Gerlach, R.~H.}, \bibinfo{author}{Chen, C.~W.}, \&
  \bibinfo{author}{Chan, N. Y.~C.} (\bibinfo{year}{2011}).
\newblock \bibinfo{title}{Bayesian time-varying quantile forecasting for
  value-at-risk in financial markets}.
\newblock {\it \bibinfo{journal}{Journal of Business \& Economic
  Statistics}\/},  {\it \bibinfo{volume}{29}\/}, \bibinfo{pages}{481--492}.
\bibitem[{Geweke \& Keane(2007)}]{KG2007}
\bibinfo{author}{Geweke, J.}, \& \bibinfo{author}{Keane, M.}
  (\bibinfo{year}{2007}).
\newblock \bibinfo{title}{Smoothly mixing regressions}.
\newblock {\it \bibinfo{journal}{Journal of Econometrics}\/},  {\it
  \bibinfo{volume}{138}\/}, \bibinfo{pages}{252--290}.
\bibitem[{Giacomini \& Komunjer(2005)}]{GK2005}
\bibinfo{author}{Giacomini, R.}, \& \bibinfo{author}{Komunjer, I.}
  (\bibinfo{year}{2005}).
\newblock \bibinfo{title}{Evaluation and combination of conditional quantile
  forecasts}.
\newblock {\it \bibinfo{journal}{Journal of Business \& Economic
  Statistics}\/},  {\it \bibinfo{volume}{23}\/}, \bibinfo{pages}{416--431}.
\bibitem[{Gneiting et~al.(2007)Gneiting, Balabdaoui \& Raftery}]{TFA2007}
\bibinfo{author}{Gneiting, T.}, \bibinfo{author}{Balabdaoui, F.}, \&
  \bibinfo{author}{Raftery, A.~E.} (\bibinfo{year}{2007}).
\newblock \bibinfo{title}{Probabilistic forecasts, calibration and sharpness}.
\newblock {\it \bibinfo{journal}{Journal of the Royal Statistical Society:
  Series B (Statistical Methodology)}\/},  {\it \bibinfo{volume}{69}\/},
  \bibinfo{pages}{243--268}.
\bibitem[{Hastings(1970)}]{H1970}
\bibinfo{author}{Hastings, W.~K.} (\bibinfo{year}{1970}).
\newblock \bibinfo{title}{Monte carlo sampling methods using markov chains and
  their applications}.
\newblock {\it \bibinfo{journal}{Biometrika}\/},  {\it \bibinfo{volume}{57}\/},
  \bibinfo{pages}{97--109}.
\bibitem[{Hoeting et~al.(1999)Hoeting, Madigan, Raftery \&
  Volinsky}]{hoeting1999}
\bibinfo{author}{Hoeting, J.~A.}, \bibinfo{author}{Madigan, D.},
  \bibinfo{author}{Raftery, A.~E.}, \& \bibinfo{author}{Volinsky, C.~T.}
  (\bibinfo{year}{1999}).
\newblock \bibinfo{title}{Bayesian model averaging: a tutorial (with
  discussion)}.
\newblock {\it \bibinfo{journal}{Statistical Science}\/},  {\it
  \bibinfo{volume}{14}\/}, \bibinfo{pages}{382--417}.
\bibitem[{Hyndman \& Athanasopoulos(2018)}]{hyndman2018}
\bibinfo{author}{Hyndman, R.~J.}, \& \bibinfo{author}{Athanasopoulos, G.}
  (\bibinfo{year}{2018}).
\newblock {\it \bibinfo{title}{Forecasting: Principles and Practice}\/}.
\newblock \bibinfo{publisher}{OTexts}.
\bibitem[{Jose \& Winkler(2008)}]{JW2008}
\bibinfo{author}{Jose, V. R.~R.}, \& \bibinfo{author}{Winkler, R.~L.}
  (\bibinfo{year}{2008}).
\newblock \bibinfo{title}{Simple robust averages of forecasts: Some empirical
  results}.
\newblock {\it \bibinfo{journal}{International Journal of Forecasting}\/},
  {\it \bibinfo{volume}{24}\/}, \bibinfo{pages}{163--169}.
\bibitem[{Kass \& Raftery(1995)}]{RA1995}
\bibinfo{author}{Kass, R.~E.}, \& \bibinfo{author}{Raftery, A.~E.}
  (\bibinfo{year}{1995}).
\newblock \bibinfo{title}{Bayes factors}.
\newblock {\it \bibinfo{journal}{Journal of the American Statistical
  Association}\/},  {\it \bibinfo{volume}{90}\/}, \bibinfo{pages}{773--795}.
\bibitem[{Kass \& Wasserman(1995)}]{KW1995}
\bibinfo{author}{Kass, R.~E.}, \& \bibinfo{author}{Wasserman, L.}
  (\bibinfo{year}{1995}).
\newblock \bibinfo{title}{A reference bayesian test for nested hypotheses and
  its relationship to the schwarz criterion}.
\newblock {\it \bibinfo{journal}{Journal of the American Statistical
  Association}\/},  {\it \bibinfo{volume}{90}\/}, \bibinfo{pages}{928--934}.
\bibitem[{Kleijn \& van~der Vaart(2006)}]{kleijn2006}
\bibinfo{author}{Kleijn, B. J.~K.}, \& \bibinfo{author}{van~der Vaart, A.~W.}
  (\bibinfo{year}{2006}).
\newblock \bibinfo{title}{Misspecification in infinite-dimensional {B}ayesian
  statistics}.
\newblock {\it \bibinfo{journal}{The Annals of Statistics}\/},  {\it
  \bibinfo{volume}{34}\/}, \bibinfo{pages}{837--877}.
\bibitem[{Koenker(2005)}]{Koenker2005}
\bibinfo{author}{Koenker, R.} (\bibinfo{year}{2005}).
\newblock {\it \bibinfo{title}{{Quantile regression}}\/}.
\newblock \bibinfo{publisher}{Cambridge University Press: Cambridge, UK}.
\bibitem[{Koenker \& Bassett~Jr(1978)}]{KB1978}
\bibinfo{author}{Koenker, R.}, \& \bibinfo{author}{Bassett~Jr, G.}
  (\bibinfo{year}{1978}).
\newblock \bibinfo{title}{Regression quantiles}.
\newblock {\it \bibinfo{journal}{Econometrica: Journal of the Econometric
  Society}\/},  (pp. \bibinfo{pages}{33--50}).
\bibitem[{Koenker \& Xiao(2006)}]{KX2006}
\bibinfo{author}{Koenker, R.}, \& \bibinfo{author}{Xiao, Z.}
  (\bibinfo{year}{2006}).
\newblock \bibinfo{title}{Quantile autoregression}.
\newblock {\it \bibinfo{journal}{Journal of the American Statistical
  Association}\/},  {\it \bibinfo{volume}{101}\/}, \bibinfo{pages}{980--990}.
\bibitem[{Li \& Dunson(2020)}]{LD2020}
\bibinfo{author}{Li, M.}, \& \bibinfo{author}{Dunson, D.~B.}
  (\bibinfo{year}{2020}).
\newblock \bibinfo{title}{Comparing and weighting imperfect models using
  d-probabilities}.
\newblock {\it \bibinfo{journal}{Journal of the American Statistical
  Association}\/},  {\it \bibinfo{volume}{115}\/}, \bibinfo{pages}{1349--1360}.
\bibitem[{Liu et~al.(2020)Liu, Li \& Morris}]{LLM2018}
\bibinfo{author}{Liu, Y.}, \bibinfo{author}{Li, M.}, \&
  \bibinfo{author}{Morris, J.~S.} (\bibinfo{year}{2020}).
\newblock \bibinfo{title}{Function-on-scalar quantile regression with
  application to mass spectrometry proteomics data}.
\newblock {\it \bibinfo{journal}{The Annals of Applied Statistics}\/},  {\it
  \bibinfo{volume}{14}\/}, \bibinfo{pages}{521--541}.
\bibitem[{Metropolis et~al.(1953)Metropolis, Rosenbluth, Rosenbluth, Teller \&
  Teller}]{MRRT1953}
\bibinfo{author}{Metropolis, N.}, \bibinfo{author}{Rosenbluth, A.~W.},
  \bibinfo{author}{Rosenbluth, M.~N.}, \bibinfo{author}{Teller, A.~H.}, \&
  \bibinfo{author}{Teller, E.} (\bibinfo{year}{1953}).
\newblock \bibinfo{title}{Equation of state calculations by fast computing
  machines}.
\newblock {\it \bibinfo{journal}{The Journal of Chemical Physics}\/},  {\it
  \bibinfo{volume}{21}\/}, \bibinfo{pages}{1087--1092}.
\bibitem[{Nakajima \& West(2013)}]{NW2013}
\bibinfo{author}{Nakajima, J.}, \& \bibinfo{author}{West, M.}
  (\bibinfo{year}{2013}).
\newblock \bibinfo{title}{Bayesian analysis of latent threshold dynamic
  models}.
\newblock {\it \bibinfo{journal}{Journal of Business \& Economic
  Statistics}\/},  {\it \bibinfo{volume}{31}\/}, \bibinfo{pages}{151--164}.
\bibitem[{Neath \& Cavanaugh(2012)}]{AJ2012}
\bibinfo{author}{Neath, A.~A.}, \& \bibinfo{author}{Cavanaugh, J.~E.}
  (\bibinfo{year}{2012}).
\newblock \bibinfo{title}{The bayesian information criterion: background,
  derivation, and applications}.
\newblock {\it \bibinfo{journal}{Wiley Interdisciplinary Reviews: Computational
  Statistics}\/},  {\it \bibinfo{volume}{4}\/}, \bibinfo{pages}{199--203}.
\bibitem[{Oka \& Qu(2011)}]{OQ2011}
\bibinfo{author}{Oka, T.}, \& \bibinfo{author}{Qu, Z.} (\bibinfo{year}{2011}).
\newblock \bibinfo{title}{Estimating structural changes in regression
  quantiles}.
\newblock {\it \bibinfo{journal}{Journal of Econometrics}\/},  {\it
  \bibinfo{volume}{162}\/}, \bibinfo{pages}{248--267}.
\bibitem[{Prado \& West(2010)}]{prado2010time}
\bibinfo{author}{Prado, R.}, \& \bibinfo{author}{West, M.}
  (\bibinfo{year}{2010}).
\newblock {\it \bibinfo{title}{Time Series: Modeling, Computation, and
  Inference}\/}.
\newblock \bibinfo{publisher}{Chapman and Hall/CRC}.
\bibitem[{Primiceri(2005)}]{primiceri2005time}
\bibinfo{author}{Primiceri, G.~E.} (\bibinfo{year}{2005}).
\newblock \bibinfo{title}{Time varying structural vector autoregressions and
  monetary policy}.
\newblock {\it \bibinfo{journal}{The Review of Economic Studies}\/},  {\it
  \bibinfo{volume}{72}\/}, \bibinfo{pages}{821--852}.
\bibitem[{Qu(2008)}]{Q2008}
\bibinfo{author}{Qu, Z.} (\bibinfo{year}{2008}).
\newblock \bibinfo{title}{Testing for structural change in regression
  quantiles}.
\newblock {\it \bibinfo{journal}{Journal of Econometrics}\/},  {\it
  \bibinfo{volume}{146}\/}, \bibinfo{pages}{170--184}.
\bibitem[{Robert(2001)}]{R2001}
\bibinfo{author}{Robert, C.~P.} (\bibinfo{year}{2001}).
\newblock {\it \bibinfo{title}{The Bayesian Choice}\/}.
\newblock \bibinfo{publisher}{Springer}.
\bibitem[{Sriram et~al.(2016)Sriram, Shi \& Ghosh}]{SK2016}
\bibinfo{author}{Sriram, K.}, \bibinfo{author}{Shi, P.}, \&
  \bibinfo{author}{Ghosh, P.} (\bibinfo{year}{2016}).
\newblock \bibinfo{title}{A bayesian quantile regression model for insurance
  company costs data}.
\newblock {\it \bibinfo{journal}{Journal of the Royal Statistical Society:
  Series A (Statistics in Society)}\/},  {\it \bibinfo{volume}{179}\/},
  \bibinfo{pages}{177--202}.
\bibitem[{{Tukey}(1974)}]{JWT1974}
\bibinfo{author}{{Tukey}, J.~W.} (\bibinfo{year}{1974}).
\newblock \bibinfo{title}{Nonlinear (nonsuperposable) methods for smoothing
  data}.
\newblock {\it \bibinfo{journal}{CONGRESS RECORD (EASCO)}\/},  (p.
  \bibinfo{pages}{673}).
\bibitem[{Yang et~al.(2016)Yang, Wang \& He}]{yang2016posterior}
\bibinfo{author}{Yang, Y.}, \bibinfo{author}{Wang, H.~J.}, \&
  \bibinfo{author}{He, X.} (\bibinfo{year}{2016}).
\newblock \bibinfo{title}{Posterior inference in bayesian quantile regression
  with asymmetric laplace likelihood}.
\newblock {\it \bibinfo{journal}{International Statistical Review}\/},  {\it
  \bibinfo{volume}{84}\/}, \bibinfo{pages}{327--344}.
\bibitem[{Yao et~al.(2018)Yao, Vehtari, Simpson \& Gelman}]{YVSG2018}
\bibinfo{author}{Yao, Y.}, \bibinfo{author}{Vehtari, A.},
  \bibinfo{author}{Simpson, D.}, \& \bibinfo{author}{Gelman, A.}
  (\bibinfo{year}{2018}).
\newblock \bibinfo{title}{Using stacking to average bayesian predictive
  distributions}.
\newblock {\it \bibinfo{journal}{Bayesian Analysis}\/},  {\it
  \bibinfo{volume}{13}\/}, \bibinfo{pages}{917--1007}.
\bibitem[{Youngman(2018)}]{BD2018}
\bibinfo{author}{Youngman, B.~D.} (\bibinfo{year}{2018}).
\newblock \bibinfo{title}{Generalized additive models for exceedances of high
  thresholds with an application to return level estimation for us wind gusts}.
\newblock {\it \bibinfo{journal}{Journal of the American Statistical
  Association}\/},  (pp. \bibinfo{pages}{1--31}).
\newblock \bibinfo{note}{To appear}.
\bibitem[{Yu \& Moyeed(2001)}]{YM2001}
\bibinfo{author}{Yu, K.}, \& \bibinfo{author}{Moyeed, R.~A.}
  (\bibinfo{year}{2001}).
\newblock \bibinfo{title}{Bayesian quantile regression}.
\newblock {\it \bibinfo{journal}{Statistics \& Probability Letters}\/},  {\it
  \bibinfo{volume}{54}\/}, \bibinfo{pages}{437--447}.

\end{thebibliography}

\end{document}